\documentclass[prd]{revtex4}

\usepackage{graphicx}
\usepackage{dcolumn}
\usepackage{bm}

\begin{document}
\renewcommand{\theequation}{\thesection.\arabic{equation}}

\title{Spin-2 Particles in Gravitational
Fields\\}

\author{G. Papini}

\affiliation{\\Department of Physics and Prairie Particle Physics
Institute,\\
 University of Regina, Regina, Sask, S4S 0A2, Canada\\}
 \affiliation{International Institute for Advanced Scientific
Studies, 89019 Vietri sul Mare (SA),
 Italy.\\}

\date{\today}

\begin{abstract}
We give a solution of the wave equation for massless, or massive
spin-$2$ particles propagating in a gravitational background. The
solution is covariant, gauge-invariant and exact to first order in
the background gravitational field. The background contribution is
confined to a phase factor from which geometrical and physical
optics can be derived. The phase also describes Mashhoon's
spin-rotation coupling and, in general, the spin-gravity
interaction.
\end{abstract}

 \maketitle

\section{\label{sec:level1}Introduction}

\setcounter{equation}{0} In a recent paper, Shen \cite{shen} has
extended Mashhoon's spin-rotation coupling \cite{mashhoon,ryder} to
include the coupling of a graviton's spin to a weak gravitomagnetic
field and has stressed the role of the self-interaction of the
gravitational field itself. Ramos and Mashhoon \cite{ramos} have
then studied the effect of the helicity-gravitomagnetic field
coupling on weak gravitational waves and shown that a gravitational
Skrotskii effect \cite{skr} exists. They further show that the
Skrotskii rotation angle is twice that expected for electromagnetic
waves.

The purpose of this paper is to extend some of the results of these
authors to include any type of weak inertial and gravitational
fields and massive as well as massless spin-2 particles and to
provide a general framework for the study of spin-$2$ particles in
external inertial and gravitational fields. This is accomplished by
solving the equation
\begin{equation}\label{spin2}
\nabla_{\alpha}\nabla^{\alpha}\Phi_{\mu\nu}+ m^2 \Phi_{\mu\nu}= 0,
\end{equation}
where $ m\geq 0$ is the mass of the particle and $ \nabla_{\alpha}$
indicates covariant differentiation. We use units $ \hbar =c=1 $.

The solution of covariant wave equations for scalar \cite{caipap1},
spin-1/2 and spin-1 particles \cite{caipap2,singh,pap1} yields in
general meaningful insights into aspects of the interaction of
quantum systems with gravity whenever the gravitational field need
not be quantized. The interaction of quantum systems with inertial
and gravitational fields produces quantum phases. Though these are
in general path-dependent, phase differences are observable, in
principle, by means of Earth bound, or space interferometers, or by
gravitational lensing.

The choice of (\ref{spin2}) requires some justification. The
propagation equations of higher spin fields contain in general
curvature dependent terms that make the formulation of these fields
particularly difficult when $ m=0$ \cite{cnock}. For spin-$2$
fields, the simplest equation of propagation used in lensing is
derived in \cite{misn} and is given by
\begin{equation}\label{MTW}
\nabla_{\alpha}\nabla^{\alpha}\Phi_{\mu\nu}+ 2
R_{\alpha\mu\beta\nu}\Phi^{\alpha\beta}=0\,.
\end{equation}
The second term in (\ref{MTW}) is localized in a region surrounding
the lens that is small relative to the distances between lens,
source and observer and is neglected when the wavelength $ \lambda$
associated with $ \Phi_{\mu\nu}$ is smaller than the typical radius
of curvature of the gravitational background \cite{tak}. For the
metric used in Sections III and IV, we find, in particular, that the
curvature term may be neglected when $ \lambda\ll
\sqrt{\frac{\rho^3}{r_g}}$, where $\rho$ is the distance from the
lens and $r_g$ its Schwarzschild radius. This condition is satisfied
in most lensing problems. It also is adequate to treat the problems
discussed in \cite{shen,ramos}.

Equation (\ref{MTW}) can be generalized to include a mass term
\begin{equation}\label{MTW1}
\nabla_{\alpha}\nabla^{\alpha}\Phi_{\mu\nu}+ 2
R_{\alpha\mu\beta\nu}\Phi^{\alpha\beta}+m^2 \Phi_{\mu\nu}=0\,.
\end{equation}
Here too the curvature term is smaller than the mass term whenever $
m>1/r_g$. For Earth bound experiments $ r_g=2 G M_\oplus/R_\oplus$
and the curvature term becomes negligible for $m>2.5\cdot 10^{-6}
GeV$. In view of the applications discussed below, the curvature
term is therefore neglected and (\ref{MTW1}) reduces to our initial
Equation (\ref{spin2}).

The background gravitational field is represented by the metric
deviation $ \gamma_{\mu\nu}=g_{\mu\nu}-\eta_{\mu\nu}$, and the
Minkowski metric $ \eta_{\mu\nu}$ has signature $-2$. To first order
in $ \gamma_{\mu\nu}$, (\ref{spin2}) can be written in the form
\begin{equation} \label{wf}
\left(\eta^{\alpha\beta}-\gamma^{\alpha\beta}\right)\partial_{\alpha}\partial_{\beta}\Phi_{\mu\nu}+
R_{\sigma\mu}\Phi_{\nu}^{\sigma}+
R_{\sigma\nu}\Phi_{\mu}^{\sigma}-2\Gamma_{\mu\alpha}^{\sigma}\partial^{\alpha}\Phi_{\nu\sigma}-
2\Gamma_{\nu\alpha}^{\sigma}
\partial^{\alpha}\Phi_{\mu\sigma}+ m^2 \Phi_{\mu\nu} = 0,
\end{equation}
where
$R_{\mu\beta}=-(1/2)\partial_{\alpha}\partial^{\alpha}\gamma_{\mu\beta}$
is the linearized Ricci tensor of the background metric and $
\Gamma_{\sigma\mu,\alpha}=1/2\left(\gamma_{\alpha\sigma,\mu}+\gamma_{\alpha\mu,\sigma}-
\gamma_{\sigma\mu,\alpha}\right)$ is the corresponding Christoffel
symbol of the first kind.

The plan of the paper is as follows. We give the solution of
(\ref{spin2}) and illustrate some of its properties in Section II.
In Section III we discuss spin-gravity coupling and geometrical
optics. Section IV is concerned with wave effects in particle
optics. In Section V we summarize and discuss the results.

\section{\label{sec:level2}Solution of the spin-2 wave equation}
\setcounter{equation}{0}

It is easy to prove, by direct substitution, that a solution of
(\ref{wf}), exact to first order in $ \gamma_{\mu\nu}$, is
represented by
\begin{eqnarray}\label{solution}
\Phi_{\mu\nu}=
\phi_{\mu\nu}-\frac{1}{4}\int_{P}^{x}dz^{\lambda}\left(\gamma_{\alpha\lambda,\beta}\left(z\right)-
\gamma_{\beta\lambda,\alpha}\left(z\right)\right)\left[\left(x^{\alpha}-z^{\alpha}\right)\partial^{\beta}\phi_{\mu\nu}
\left(x\right)-\left(x^{\beta}-z^{\beta}\right)\partial^{\alpha}\phi_{\mu\nu}\left(x\right)\right]
\\ \nonumber + \frac{1}{2}
\int_{P}^{x}dz^{\lambda}\gamma_{\alpha\lambda}\left(z\right)\partial^{\alpha}\phi_{\mu\nu}\left(x\right)+
\int_{P}^{x}dz^{\lambda}\Gamma_{\mu\lambda,\sigma}\left(z\right)\phi_{\nu}^{\sigma}\left(x\right)+
\int_{P}^{x}dz^{\lambda}\Gamma_{\nu\lambda,\sigma}\left(z\right)
\phi_{\mu}^{\sigma}\left(x\right),
\end{eqnarray}
where $\phi_{\mu\nu}$ satisfies the field-free equation
\begin{equation} \label{ff}
\left(\partial_{\alpha}\partial^{\alpha}+
m^2\right)\phi_{\mu\nu}\left(x\right)=0 \,,
\end{equation}
and the Lanczos-DeDonder gauge condition $
\gamma_{\alpha\nu,}^{\,\,\,\,\
  \nu}-\frac{1}{2}\gamma_{\sigma,\alpha}^\sigma = 0$  has been
  used. In (\ref{solution}) $ P$ is a fixed reference point and $ x $ a generic point
along the particle's worldline.

 The particular case $ m=0$ yields
the solution $ \Phi_{\mu\nu}$ for a linearized gravitational field $
\phi_{\mu\nu}$ propagating in a background gravitational field $
\gamma_{\mu\nu}$.

The solution (\ref{solution}) applies equally well when $
\phi_{\mu\nu}$ is a plane wave or a wave packet solution of
(\ref{ff}). No additional approximations are made regarding $
\Phi_{\mu\nu}$ that obviously satisfies the equation (\ref{ff}) when
$ \gamma_{\mu\nu}$ vanishes.

We show below that, as in \cite{caipap1,caipap2}, the solution is
manifestly covariant. It also is completely gauge invariant and the
effect of gravitation is entirely contained in the phase of the wave
function. In fact, (\ref{solution}) can be written in the form $
\Phi_{\mu\nu}=exp(-i\xi) \phi_{\mu\nu}\simeq (1-i\xi)\phi_{\mu\nu}$
or, explicitly, as
\begin{eqnarray}\label{sol'}
\Phi_{\mu\nu}\left(x\right)=\phi_{\mu\nu}\left(x\right)+\frac{1}{2}\int_{P}^{x}dz^{\lambda}\gamma_{\alpha\lambda}
\left(z\right)
\partial^{\alpha}\phi_{\mu\nu}\left(x\right)-
\frac{1}{2}\int_{P}^{x}dz^{\lambda}\left(\gamma_{\alpha\lambda,\beta}\left(z\right)-\gamma_{\beta\lambda,\alpha}
\left(z\right)\right)\left[\left(x^{\alpha}-z^{\alpha}\right)\partial^{\beta}+iS^{\alpha\beta}\right]\phi_{\mu\nu}
\left(x\right)\\\nonumber - \frac{i}{2}
\int_{P}^{x}dz^{\lambda}\gamma_{\beta\sigma,\lambda}\left(z\right)T^{\beta\sigma}\phi_{\mu\nu}\left(x\right)\,,
\end{eqnarray}
where \begin{eqnarray}\label{ST}
S^{\alpha\beta}\phi_{\mu\nu}&\equiv& \frac{i}{2}
\left(\delta_{\sigma}^{\alpha}\delta_{\mu}^{\beta}\delta_{\nu}^{\tau}-\delta_{\sigma}^{\beta}
\delta_{\mu}^{\alpha}\delta_{\nu}^{\tau}+\delta_{\sigma}^{\alpha}\delta_{\nu}^{\beta}\delta_{\mu}^{\tau}-
\delta_{\sigma}^{\beta}\delta_{\nu}^{\alpha}\delta_{\mu}^{\tau}\right)\phi_{\tau}^{\sigma}
\\ \nonumber
T^{\beta\sigma}\phi_{\mu\nu}&\equiv&
i\left(\delta_{\mu}^{\beta}\delta_{\nu}^{\tau}+\delta_{\nu}^{\beta}\delta_{\mu}^{\tau}\right)\phi^{\sigma}_{\tau}\,.
\end{eqnarray}
From $ S^{\alpha\beta}$ one constructs the rotation matrices $
S_{i}=-2i \epsilon_{ijk}S^{jk}$ that satisfy the commutation
relations $ [S_{i},S_{j}]=i \epsilon_{ijk}S_{k}$. The spin-gravity
interaction is therefore contained in the term
\begin{equation}\label{SG}
\Phi_{\mu\nu}'\equiv
-\frac{i}{2}\int_{P}^{x}dz^{\lambda}\left(\gamma_{\alpha\lambda,\beta}-\gamma_{\beta\lambda,\alpha}\right)
S^{\alpha\beta}\phi_{\mu\nu}\left(x\right)=\frac{1}{2}\int_{P}^{x}dz^{\lambda}\left[\left(\gamma_{\sigma\lambda,\mu}-
\gamma_{\mu\lambda,\sigma}\right)\phi_{\nu}^{\sigma}+\left(\gamma_{\sigma\lambda,\nu}-\gamma_{\nu\lambda,\sigma}\right)
\phi_{\mu}^{\sigma}\right].
\end{equation}
The solution (\ref{solution}) is invariant under the gauge
transformations $ \gamma_{\mu\nu}\rightarrow
\gamma_{\mu\nu}-\xi_{\mu,\nu}-\xi_{\nu,\mu}$, where $ \xi_{\mu}$ are
small quantities of the first order. If, in fact, we choose a closed
integration path $ \Gamma $, Stokes theorem transforms the first
three integrals of (\ref{sol'}) into $ 1/4
\int_{\Sigma}d\sigma^{\lambda\kappa}R_{\lambda\kappa\alpha\beta}\left(L^{\alpha\beta}+S^{\alpha\beta}\right)
\phi_{\mu\nu}$, where $ \Sigma$ is the surface bound by $ \Gamma$,
 $ J^{\alpha\beta}=L^{\alpha\beta}+S^{\alpha\beta}$ is the total
angular momentum of the particle and $
R_{\lambda\kappa\alpha\beta}=1/2\left(\gamma_{\lambda\beta,\kappa\alpha}+\gamma_{\kappa\alpha,\lambda\beta}-
\gamma_{\lambda\alpha,\kappa\beta}-\gamma_{\kappa\beta,\lambda\alpha}\right)$
is the linearized Riemann tensor. For the same path $ \Gamma $ the
integral involving $ T^{\beta\sigma}$ in (\ref{sol'}) vanishes. It
behaves like a gauge term and may therefore be dropped. For the same
closed paths, (\ref{sol'}) gives
\begin{equation}\label{sol"}
\Phi_{\mu\nu}\simeq \left(1-i\xi\right)\phi_{\mu\nu}= \left(1 -
\frac{i}{4}\int_{\Sigma}d\sigma^{\lambda\kappa}R_{\lambda\kappa\alpha\beta}J^{\alpha\beta}\right)\phi_{\mu\nu}\,,
\end{equation}
which obviously is covariant and gauge invariant. For practical
applications (\ref{solution}) is easier to use.

The phase $ \xi$ is sometimes referred to as gravitational Berry
phase\cite{berry} because space-time plays in it the role of Berry's
parameter space\cite{caipap3}.

\section{\label{sec:level3}Helicity-gravity coupling and geometrical optics}
\setcounter{equation}{0}

The helicity-rotation coupling for massless, or massive spin-$2$
particles follows immediately from the $S^{\alpha\beta}$ term in
(\ref{sol'}). In fact, the particle energy is changed by virtue of
its spin by an amount given by the time integral of this spin term
\begin{equation}\label{hrc}
\xi^{hr}=-\frac{1}{2}\int_{P}^{x}dz^{0}\left(\gamma_{\alpha0,\beta}-\gamma_{\beta0,\alpha}\right)S^{\alpha\beta}\,,
\end{equation}
that must then be applied to a solution of (\ref{ff}). For rotation
about the $ x^{3}$-axis, $ \gamma_{0i}=\Omega(y,-x,0) $, we find $
\xi^{hr}=-\int_P^x dz^0 2 \Omega S^3 $ and the energy of the
particle therefore changes by $\pm 2\Omega$, where the factor $\pm
2$ refers to the particle's helicity, as discussed by Ramos and
Mashhoon \cite{ramos}. Equation (\ref{hrc}) extends their result to
any weak gravitational, or inertial field.

The effect of (\ref{SG}) on $ \phi_{\mu\nu}$ can be easily seen in
the case of a gravitational wave propagating in the $x$-direction
and represented by the components $
\phi_{22}=-\phi_{33}=\varepsilon_{22}exp\left[ik\left(t-x\right)\right]$
and $ \phi_{23}=\varepsilon_{23}exp\left[ik\left(t-x\right)\right]$.
For an observer rotating about the $ x$-axis the metric is $
\gamma_{00}=-\Omega^2 r^2\,,
\gamma_{11}=\gamma_{22}=\gamma_{33}=-1\,,\gamma_{0i}=\Omega(0,z,-y)
$. Then the two independent polarizations $ \phi_{23}$ and $
\phi_{22}-\phi_{33} $ are transformed by $ S_{\alpha\beta}$ into $
\Phi_{23}=-2\,\Omega\,
\left(x^{0}-x_P^0\right)/2\,\left(\phi_{22}-\phi_{33}\right)$ and $
1/2\,\left(\Phi_{22}-\Phi_{33}\right)=2\,\Omega(x^0-x_P^0)\,\phi_{23}$.

For closed integration paths and vanishing spin, (\ref{sol'})
coincides with the solution of a scalar particle in a gravitational
field, as expected. This proves the frequently quoted statement
\cite{thorne} that gravitational radiation propagating in a
gravitational background is affected by gravitation in the same way
that electromagnetic radiation is (when the photon spin is
neglected).

The geometrical optics approximation follows immediately from
(\ref{sol'}) with $ S_{\alpha\beta}=0\,, T_{\alpha\beta}=0$. We
obtain from (\ref{solution}) and (\ref{sol'})
\begin{equation}\label{phi0}
\Phi_{\mu\nu}^{0}=\phi_{\mu\nu}\left(x\right)+\frac{1}{2}\int_{P}^{x}dz^{\lambda}\gamma_{\alpha\lambda}\left(z\right)
\partial^{\alpha}\phi_{\mu\nu}\left(x\right)-
\frac{1}{2}\int_{P}^{x}dz^{\lambda}\left(\gamma_{\alpha\lambda,\beta}\left(z\right)-\gamma_{\beta\lambda,\alpha}
\left(z\right)\right)\left[\left(x^{\alpha}-z^{\alpha}\right)\partial^{\beta}\right]\phi_{\mu\nu}\left(x\right)\\.
\end{equation}
We first calculate the general relativistic deflection of a spin-2
particle in a gravitational field. It follows immediately from $
\Phi_{\mu\nu}^{0}$. Assuming, for simplicity, that the spin-2
particles are massless and propagate along the $z$-direction, so
that $k^\alpha\simeq (k, 0, 0, k)$, and $ds^2=0$ or $dt=dz$, using
plane waves for $\phi_{\mu\nu}$ and writing
 \begin{equation}\label{CHI}
 \chi = k_{\sigma}x^{\sigma}-\frac{1}{4}\int_P^x dz^\lambda
 (\gamma_{\alpha\lambda,\beta}(z)-\gamma_{\beta\lambda,
 \alpha}(z))[(x^\alpha-z^\alpha)k^\beta-(x^\beta-z^\beta)k^\alpha]+
 \frac{1}{2}\int_P^x dz^\lambda \gamma_{\alpha\lambda}(z)
 k^\alpha
 \,,
 \end{equation}
we obtain the particle momentum
\begin{equation}\label{momentum}
 \tilde{k}_\sigma = \frac{\partial \chi}{\partial x^\sigma}\equiv \chi_{,\sigma}=k_{\sigma}-\frac{1}{2}\int_P^x dz^\lambda
\left(\gamma_{\sigma\lambda,\beta}-
\gamma_{\beta\lambda,\sigma}\right)k^\beta+\frac{1}{2}\gamma_{\alpha\sigma}k^\alpha\,.
\end{equation}
It is easy to show from  (\ref{momentum}) that $\chi $ satisfies the
eikonal equation $ g^{\alpha\beta} \chi_{,\alpha} \chi_{,\beta}=0 $.

The calculation of the deflection angle is particularly simple if we
choose the background metric
\begin{equation}\label{metric}
\gamma_{00}=2U(\rho)\,\,,\, \gamma_{ij}=2U(\rho)\delta_{ij}\,,
\end{equation}
where $U(\rho)=-GM/\rho$ and $\rho=\sqrt{x^2+y^2+z^2}$, which is
frequently used in gravitational lensing. For this metric, $\chi$ is
given by
\begin{equation}\label{chi}
 \chi  \simeq  -\frac{k}{2}\int_P^x \left[
   (x-x')\phi_{, z'}dx'+(y-y')\phi_{,\, z'}dy'-2[(x-x')\phi_{,\, x'}+(y-y')\phi_{,
   \, y'}]dz'\right]+k\int_P^x dz' \phi
   \,.
\end{equation}
 The space components of the momentum are therefore
\begin{eqnarray}\label{k1}
 \tilde{k}_1 &=& 2k\int_P^x  \left(-\frac{1}{2}\frac{\partial U}{\partial z}\,dx
 +\frac{\partial U}{\partial x}dz\right)
 \,,\label{k2}\\
 \tilde{k}_2 &=& 2k\int_P^x  \left(-\frac{1}{2}\frac{\partial U}{\partial z}\,dy  +
 \frac{\partial U}{\partial y}dz\right) \,, \\
  \tilde{k}_3 &=& k(1+U)\,. \label{k3}
\end{eqnarray}
We then have
\begin{equation}\label{kvector}
  {\bf \tilde{k}}={\bf \tilde{k}}_\perp +k_3\, {\bf e}_3\,, \quad {\bf
  \tilde{k}}_\perp=k_1\, {\bf e}_1+k_1\, {\bf e}_2\,,
\end{equation}
where ${\bf \tilde{k}}_\perp$ is the component of the momentum
orthogonal to the direction of propagation of the particles.

Since only phase differences are physical, it is convenient to
choose the space-time path by placing the particle source at a
distance very large relative to the dimensions of $ M $, while the
generic point is located at $ z$ along the $z$ direction and $z\gg
x, y$. Equations (\ref{k1})-(\ref{k3}) simplify to
\begin{eqnarray}\label{k1bis}
 \tilde{k}_1 &=& 2k\int_{-\infty}^z  \frac{\partial U}{\partial x}dz = k\, \frac{2GM}{R^2}\, x\left(1+\frac{z}{r}\right)\,,\\
 \tilde{k}_2 &=& 2k\int_{-\infty}^z  \frac{\partial U}{\partial y}dz=k\, \frac{2GM}{R^2}\, y\left(1+\frac{z}{r}\right)\,, \label{k2bis} \\
 \tilde{k}_3 &=& k(1+U)\,, \label{k3bis}
\end{eqnarray}
where $R=\sqrt{x^2+y^2}$. By defining the deflection angle as
\begin{equation}\label{theta}
  \tan \theta = \frac{\tilde{k}_\perp}{\tilde{k}_3}\,,
\end{equation}
we find
\begin{equation}\label{thetak}
  \tan \theta \sim \theta \sim \frac{2GM}{R}\left(1+\frac{z}{r}\right)\,,
\end{equation}
and, in the limit $z\to \infty$, we obtain the usual Einstein result
\begin{equation}\label{thetaM}
  \theta_M\sim \frac{4GM}{R}\,.
\end{equation}
The index of refraction can be derived from the known equation $ n =
\tilde{k}/\tilde{k}_0 $. Choosing the direction of propagation of
the particle along the $x^3-$axis, and using (\ref{momentum}), we
find
\begin{equation}\label{n}
n\simeq 1 +
\frac{1}{k_0}\left(\chi_{,3}-\chi_{,0}\right)-\frac{m^2}{2k_0^2}\left(1-\frac{1}{k_0}\chi_{,0}\right)
\end{equation}
and, again, for $ k_0\gg m $, or for vanishing $ m $,
\begin{equation}\label{n'}
n\simeq 1 + \frac{1}{2k_0}\left[-\int_P^x dz^\lambda
\left(\gamma_{3\lambda,\beta}-\gamma_{\beta\lambda,3}\right)k^{\beta}+\gamma_{\alpha3}k^\alpha
+\int_P^x dz^\lambda
\left(\gamma_{0\lambda,\beta}-\gamma_{\beta\lambda,0}\right)k^\beta
-\gamma_{\alpha0}k^\alpha\right]\,.
\end{equation}
In the case of the metric (\ref{metric}), we obtain
\begin{equation}\label{n"}
n\simeq 1 + \int_P^x dz^0 \gamma_{00,3}=1-\frac{2GM}{r}\,,
\end{equation}
which is a known result.

\section{\label{sec:level4}Wave optics}
\setcounter{equation}{0}

The applications of (\ref{phi0}) to interferometry given
in\cite{caipap3,cai} cover a variety of metrics and physical
situations, from the study of rotation and Earth's field to the
detection of gravitational radiation and the calculation of effects
due to a Lense-Thirring background. The same equation and some of
its generalizations can also be applied to the study of quantum
fluids and Boson condensates. Here we apply (\ref{phi0}) to the
investigation of wave effects in lensing.

As an example, we consider the propagation of gravitons, or extreme
relativistic spin-2 particles in the background metric
(\ref{metric}). Wave optics effects can best be seen by considering
a double slit experiment, or alternatively, the lensing
configuration illustrated in Fig.\ref{fig:Lensing}. For simplicity,
we use a planar arrangement in which particles, source,
gravitational deflector and observer lie in the same plane. The
particles are emitted at $S$ and interfere at $O$, where the
observer is located, following the paths $SLO$ and $SPO$. The
interference and diffraction effects depend on the phase difference
experienced by the particles along the different paths and on the
gravitational background generated by the spherically symmetric lens
at $M$. We also use a plane wave solution of (\ref{ff}) of the form
$\phi_{\mu\nu} = e^{-ik_\sigma x^\sigma}\epsilon_{\mu\nu}$ and
assume, for simplicity, that $k^1 =0$, so that propagation is
entirely in the $(x^2, x^3$)-plane. In this planar set-up $
\gamma_{11}$ plays no role. The corresponding wave amplitude is
therefore
\begin{eqnarray}
 \Phi_{\mu\nu}^{0} &=& -ie^{-i k_\sigma x^\sigma}\left\{
 1-\frac{1}{2}\left[\int dz^0 \gamma_{00,2} (x^0-z^0) k^2 + \int
 dz^0 \gamma_{00, 3}(x^0-z^0) k^3-\int dz^0 \gamma_{00,2}
 (x^2-z^2)k^0
 \right.\right. \nonumber \\
 & - &\int dz^0 \gamma_{00, 3}(x^3-z^3) k^0 + \int dz^2 \gamma_{22,3}(x^2-z^2)k^3+
 \int dz^3 \gamma_{33,2}(x^3-z^3)
 k^2-\int dz^2 \gamma_{22,3}
 (x^3-z^3)k^2  \nonumber\\
  & - & \left. \left. \int dz^3 \gamma_{33,2}(x^2-z^2) k^3\right] +
  \frac{1}{2}\left[\int dz^0 \gamma_{00}
 k^0+\int dz^2 \gamma_{22} k^2+\int dz^{3}\gamma_{33}k^{3}\right] \right\}\,. \label{*}
\end{eqnarray}
The phase must now be calculated along the different paths $SP+PO$
and $SL+LO$, taking into account the values of $k^i$ in the various
intervals. The phase difference is therefore given by $\Delta
\tilde{\phi} = \tilde{\phi}_{SLO}-\tilde{\phi}_{SPO}$.

It is convenient to transform all space integrations into
integrations over $z^0$. Along $SL$ we have
\begin{equation}\label{2*}
  U= \frac{-GM}{q_{SL}(z^0)^{1/2}}\,, \quad q_{SL}(z^0)\equiv
  (r_L-z^0)^2+b^{+\, 2}+2(r_L-z^0)b^+ \cos \varphi^+\,,
\end{equation}
 and
 $ k^2 = k \cos \varphi^+, \quad k^3= k\sin \varphi^+$
 and $r_L \sin \varphi^+ =
D_{dS}$. We find
\begin{eqnarray}\label{3*}
 -\frac{\Delta \tilde{\phi}_{SL}}{GM} &=& 2\int_0^{r_L} dz^0
 q_{SL}(z^0)^{-3/2}(z^0-r_L-b^+
 \cos\varphi^+)(r_L-z^0)[k^2\frac{\sin\varphi^{+\,2}}{\cos\varphi^{+}}+k^{3}\frac{\cos\varphi^{+\,2}}
 {\sin\varphi^{+}}+k]
 \\
  && +
 \int_0^{r_L}dz^0 q_{SL}(z^0)^{-1/2}[- k -
 k^2 \cos\varphi^{+} -k^3 \sin\varphi^{+}]\,. \nonumber
\end{eqnarray}
Analogously, for $LO$ we have
\begin{equation}\label{4*}
  U= -\frac{GM}{q_{LO}(z^0)^{1/2}}\,, \quad q_{LO}(z^0)\equiv
  (R_1-z^0+r_L)^2 + r_0^2 - 2(R_1-z^0+r_L)r_0\cos \theta^+ \,,
\end{equation}
 while
 $ k^2 = k \sin \theta^+, \quad k^3= k\cos \theta^+, \quad R_{1}=\sqrt{r_0^{2}+ b^{+2}}
 $,
 and the change in phase is
\begin{eqnarray}\label{5*}
 -\frac{\Delta \tilde{\phi}_{LO}}{GM} &=& 2\int_{r_L}^{r_L+R_1} dz^0
 q_{LO}(z^0)^{-3/2}(R_1-z^0+r_L+r_0
 \cos\theta^+)(r_L+R_1-z^0)[k^2\frac{\cos\theta^{+\,2}}{\sin\theta^+}+k^3
 \frac{\sin\theta^{+\,2}}{\cos\theta^+}-k]
 \\
  && + \int_{r_L}^{r_L+R_1}dz^0 q_{LO}^{-1/2}[-k -k^2\sin \theta^+
 -k^3\cos \theta^+]\,. \nonumber
\end{eqnarray}
For $SP$ we find
\begin{equation}\label{6*}
  U= -\frac{GM}{q_{SP}(z^0)^{1/2}}\,, \quad q_{SP}(z^0)\equiv
  b^{-\, 2}+(R-z^0)^2-2(R-z^0)b^- \cos \gamma\,,
\end{equation}
 $ k^2 = k \cos \gamma, \quad k^3= k\sin \gamma\,, \quad
 \tan\gamma=D_{dS}/(s+b^{-})\,, \quad
 R=\sqrt{D_{dS}^2+(s+b^-)^2}$,
 and the corresponding change in phase is
\begin{eqnarray}\label{7*}
 -\frac{\Delta \tilde{\phi}_{SP}}{GM} &=& 2\int_0^{R} dz^0
 q_{SP}(z^0)^{-3/2}(R-z^0- b^- \cos\gamma)(R-z^0)\left[k^2\frac{\sin\gamma^{2}}{\cos\gamma}
 +k^3\frac{\cos\gamma^{2}}{\sin\gamma} + k\right]
 \\
 && +
 \int_0^{R}dz^0 q_{SP}(z^0)^{-1/2}[-k -k^2\cos \gamma
 - k^3\sin \gamma]\,. \nonumber
\end{eqnarray}
Finally, for $PO$ we get
\begin{equation}\label{8*}
  U= -\frac{GM}{q_{PO}(z^0)^{1/2}}\,, \quad q_{PO}(z^0)\equiv
  r_0^2 + (R_2+R-z^0)^2 - 2(R_2+R-z^0)r_0\cos \theta^- \,,
\end{equation}
  $ k^2 =- k \sin \theta^-, \quad k^3= k\cos \theta^-\,,
 \quad R_2=\sqrt{r_0^2+b^{-\, 2}}$,
 and the relative change in phase is
\begin{eqnarray}\label{9*}
 -\frac{\Delta \tilde{\phi}_{PO}}{GM} &=& 2\int_{R}^{R+R_2} dz^0
 q_{PO}(z^0)^{-3/2}(z^0-R_2-R+r_0
 \cos\theta^-)(R+R_2-z^0)\left[-k^2\frac{\cos\theta^{-\,2}}{\sin\theta^-}+
 k^3\frac{\sin\theta^{-\,2}}{\cos\theta^-}-k\right]
 \\
  && + \int_{R}^{R+R_2}dz^0 q_{PO}^{-1/2}[-k +k^2\sin \theta^- -k^3\cos \theta^-]\,. \nonumber
\end{eqnarray}
The total change in phase is therefore
 $\Delta \tilde{\phi}= \Delta \tilde{\phi}_{SL} +\Delta
\tilde{\phi}_{LO}-\Delta \tilde{\phi}_{SP}-\Delta
\tilde{\phi}_{PO}$. All integrations in (\ref{3*}), (\ref{5*}).
(\ref{7*}) and (\ref{9*}) can be performed exactly.
 All results can be expressed in terms of physical variables
 $r_s$, $r_0$, $b^+$, $b^-$, and $s$ and lensing variables $D_s$,
 $D_{ds}$, $D_d$, $\theta^+$, $\theta^-$, and $\beta$. The final
 result is
\begin{eqnarray}\label{FR}
\Delta\tilde{\phi} & = & \tilde{y} \left\{\ln \left(-\sqrt{D_{dS}^2
+\left(s + b^-\right)^2} + b^- \cos\gamma + r_S \right)- \ln
\left(b^- \left(1+ \cos\gamma\right)\right)
\right.\nonumber \\
& + & \ln \left(b^+ \left(1-\cos\varphi^+ \right)\right)-
\ln\left(r_S -r_L-b^+\cos\varphi^+\right)
\nonumber \\
& + & \ln \left( b^- + r_0 \cos\theta^- -\sqrt{b^{-\,2}+
r_0^2}\right)-\ln \left(r_0\left(1+ \cos\theta^-\right)\right)
\nonumber \\
& + & \left.\ln\left(r_0\left(1+ \cos\theta^+\right)\right)-
\ln\left(b^+ +r_0 \cos\theta^+ -\sqrt{b^{+\,2}+r_0^2}\right)
\right\} \,,
\end{eqnarray}
where $ r_S^2 = b^{+\,2}+r_L^2+2b^+r_L\cos\varphi^+ \,,
r_L^2=D_{dS}^2 +(s-b^+)^2 \,,  \varphi^+ +\alpha^+ +\alpha^- +\gamma
-\theta^+ -\theta^- =\pi $ and $ \tilde{y}= 2GMk$. This result is
exact and independent of the value of $\tilde{y} $.

A simple quantum mechanical calculation indicates that the
probability of finding particles at $ O $ contains an oscillating
term (two-image interference) that is proportional to $ \cos^2
\Delta\tilde{\phi}/2$. In the particular case $ b^+ \sim b^- \equiv
b\,, \varphi^+\sim \gamma \,, r_S\sim r_L\gg (b,s) \,,\theta^+\sim
\theta^-\equiv \theta\,,\alpha^+\sim \alpha^-\equiv \alpha$, we
obtain from (\ref{FR}), the expression
\begin{equation}\label{diff}
\Delta\tilde{\phi} \sim \tilde{y}\ln \frac{r_L
\left(1+\sin\left(\theta -\alpha\right)\right)}{b\sin\left(\theta
-\alpha \right)} \,,
\end{equation}
which is approximate to terms of $ O(b/r_L) $ and higher in the
argument of the logarithm. The overall probability $ P_0$ of finding
particles at $ O$ is therefore
\begin{equation}\label{prob}
P_0 \propto \cos^2\left\{\frac{\tilde{y}}{2} \ln\left[\frac{r_L}{2b}
\frac{\left(1+\tan\frac{\theta -\alpha}{2}\right)^2}
{\tan\frac{\theta -\alpha}{2} }\right] \right\}\,,
\end{equation}
which exhibits an oscillating behavior typical of combined
interference and diffraction effects. Higher order terms in $ b/r_L
$ would in general prevent the logarithmic term from diverging when
$ \theta \rightarrow \alpha $.

\section{\label{sec:level5}Conclusions}
\setcounter{equation}{0}

We have solved the wave equation (\ref{spin2}) for massless and
massive spin-$ 2$ particles propagating in a background
gravitational field. The solution is exact to first order in $
\gamma_{\mu\nu}$, is covariant and gauge-invariant and is known
whenever a solution $ \phi_{\mu\nu}$ of the free wave equation is
known.

The external gravitational field, represented by the background
metric, only appears in the phase of the wave function. It is
precisely this phase that provides the general framework for the
study of spin-$2$ particles.

The spin-gravity coupling and Mashhoon's helicity-rotation
interaction follow from the gravity-induced phase, (\ref{SG}). The
origin of (\ref{SG}) resides in the skew-symmetric part of the
space-time connection terms in (\ref{solution}), while, in the case
of fermions, it is the spinorial connection \cite{caipap2} that
accounts for $S_{\alpha\beta}$. The spin term $ S_{\alpha\beta} $
affects the polarization of a gravitational wave, as shown in
Section III. It also plays a role in the collision of two
gravitational waves. If these are represented by $
\phi_{22}=-\phi_{33}=\varepsilon_{22}exp\left[ik\left
(t-x\right)\right]\,, \phi_{23}=\varepsilon_{23}exp\left[ik\left
(t-x\right)\right]$ and the gravitational background is a wave of
the same polarization, but proceeding along the negative direction
of the $x$-axis, then the corresponding metric is
\begin{equation}\label{met}
ds^2=2 \Phi_{02}' dx^0 dx^2 +2\Phi_{03}'dx^0 dx^3 +2\Phi_{12}'dx^1
dx^2 + 2\Phi_{13}dx^1 dx^3 +2\left(\Phi_{23}^0+\Phi_{23}'\right)dx^2
dx^3\,,
\end{equation}
where
\begin{eqnarray}\label{ms}
\Phi_{02}'&=&
\frac{-ik}{2}\left[\left(\gamma_{22}\phi_{22}+\gamma_{32}\phi_{23}\right)x^2+
\left(\gamma_{23}\phi_{23}+\gamma_{33}\phi_{23}\right)x^3\right]\,;
\\ \nonumber
\Phi_{03}'&=&\frac{-ik}{2}\left[\left(\gamma_{22}\phi_{32}+\gamma_{32}\phi_{23}\right)x^2+
\left(\gamma_{23}\phi_{32}+\gamma_{33}\phi_{33}\right)x^3\right]\,;
\\ \nonumber
\Phi_{12}'&=&
\frac{-ik}{2}\left[\left(\gamma_{22}\phi_{22}+\gamma_{32}\phi_{23}\right)x^2+\left(\gamma_{23}\phi_{22}+
\gamma_{33}\phi_{23}\right)x^3\right]\,;
\\
\nonumber
\Phi_{13}'&=&\frac{-ik}{2}\left[\left(\gamma_{22}\phi_{23}+
\gamma_{32}\phi_{33}\right)x^2+\left(\gamma_{23}\phi_{32}+\gamma_{33}\phi_{33}\right)x^3\right]\,;
\\ \nonumber
\Phi_{23}'&=&\phi_{22}\left(\gamma_{32}- \gamma_{33}\right)\,,
\end{eqnarray}
and the collision takes place at the origin of the coordinates. The
metric (\ref{met}) has a singularity at $ x^2=x^3=0 $. More complete
treatments of this problem show that this is a curvature singularity
\cite{szek1,penr,yurts,grif}.

From the phase we have derived the geometrical optics of the
particles and verified that their deflection is that predicted by
Einstein. The gravitational background behaves as a material medium
of index of refraction $ n$ given by (\ref{n}).

\begin{figure}
\includegraphics{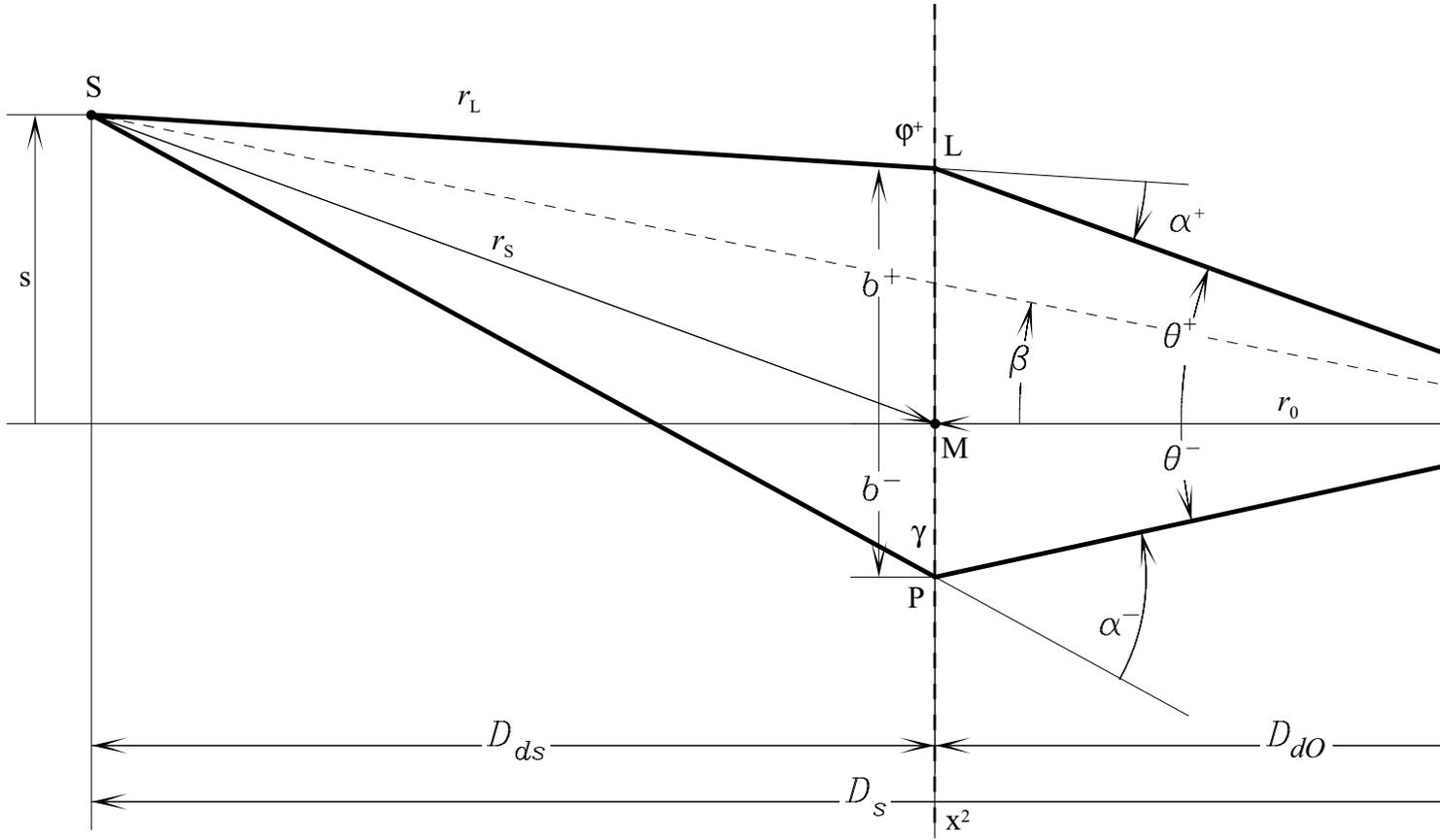}
\caption{\label{fig:Lensing} Geometry of a two-image gravitational
lens or, equivalently, of a double slit interference experiment. The
solid lines represent the particle paths between the particle source
at $S$ and the observer at $O$. $M$ is the spherically symmetric
gravitational lens. $S, M, O$ and the particle paths lie in the same
plane. The physical variables are $ r_{S}, r_{0}, b^{\pm}, s$, while
the lensing variables are indicated by $ D_{dS}, D_{dO}, D_{S},
\theta^{\pm}, \beta$. $ \alpha^{\pm}$ are the deflection angles.}
\end{figure}

Wave optics too can be extracted from the phase. We have derived an
exact expression for the phase change $ \Delta\tilde{\phi}$ given by
(\ref{FR}) and have shown that (\ref{FR}) represents interference
and diffraction effects. In gravitational lensing
\cite{ehlers,deguchi} and in the gravitational lensing of
gravitational waves \cite{tak}, wave effects for a point source
depend on the parameter $ \tilde{y}$ which gives an indication of
the maximum magnification of the wave flux, or, alternatively, of
the number of Fresnel zones contributing to lensing. Different
values of $ \tilde{y}$ require, in general, different
approximations, or different solutions of the wave equation. In
particular, diffraction effects are expected to be considerable when
$ \tilde{y}\simeq 1$. In our approach, (\ref{FR}) holds true
regardless of the value of $ \tilde{y}$. Wave optics problems
usually deal with spherical wave solutions of Helmholtz equation in
which gravity appears in the form of a potential. The extension of
our findings to include spherical wave solutions is, of course,
allowed by (\ref{ff}), but results in additional terms in
(\ref{phi0}) and in a more cumbersome, but still exact final result.
It is left to a future, specific application in which the planar
configuration of Fig.\ref{fig:Lensing} will be rescinded. In the
present approach, however, gravity makes itself felt in a rather
more subtle way than just through a single potential, as evidenced
by (\ref{wf}).

The framework developed can also be used in the interferometry of
atoms and molecules. A laboratory instrument capable of using
coherent beams of atoms or molecules would go a long way in probing
the interface between gravitational theories and quantum mechanics.
For instance, the phase shift of a particle beam in the
Lense-Thirring field of the Earth is \cite{pap}
\begin{equation}\label{LT}
\Delta \tilde{\phi}_{LT} = \frac{4G}{R^{3}_{\oplus}} J_{\oplus}
m\ell^2 \,,
\end{equation} where $J_{\oplus} =2M_{\oplus}R^{2}_{\oplus}\Omega
/5 $ is the angular momentum of Earth (assumed spherical and
homogeneous), $R_{\oplus}$ its radius and $\ell$ the typical
dimension of the interferometer. For neutron interferometers with $
\ell\sim10^2cm$, we find $ \Delta \tilde{\phi}_{LT}\sim10^{-7}rad $.
The value of the phase difference increases with $ m$ and $ \ell^2$.
This suggests that the development and use of large, heavy particle
interferometers would be particularly advantageous in attempts to
measure gravitational effects. When (\ref{FR}) and (\ref{metric})
are used in the case of a square interferometer and extreme
relativistic particles, we however obtain $ \tilde{\phi}\simeq GMk$,
irrespective of the size of the interferometer. This is as expected.
In fact, for the particular configuration of Fig.\ref{fig:Lensing}
(and unlike the problems considered in \cite{caipap1,cai}), the
gravitational flux of the source is completely contained in the
integration path $ SLOPS$ and $ \Delta\tilde{\phi}$ can not be made
larger by increasing the dimensions of the interferometer.

\begin{acknowledgments}
The author wishes to thank E. di Marino and G. Lambiase for their
help in preparing the figure.
\end{acknowledgments}

\end{document}